\newcommand{\be}{\begin{equation}}
\newcommand{\ee}{\end{equation}}
\newcommand{\bea}{\begin{eqnarray}}
\newcommand{\eea}{\end{eqnarray}}
\newcommand{\ba}[1]{\begin{array}{#1}}
\newcommand{\ea}{\end{array}}

\newcommand{\diracslash}[1]{#1\llap{/\kern2pt}}

\documentclass[aps,pra,preprint,showpacs,amsmath,amssymb]{revtex4}
\usepackage{epsfig}
\usepackage{amssymb}
\usepackage{times}
\usepackage{amssymb}
\usepackage{amsmath}
\usepackage{mathrsfs}
\usepackage{graphicx}
\usepackage{epsfig}
\usepackage{dcolumn}
\usepackage{color}
\usepackage{bm}
\begin{document}
\title{ Creation and manipulation of bound states in continuum with lasers: Applications 
to cold atoms and molecules}
\author{Bimalendu Deb$^1$ and G. S. Agarwal$^2$ }
\affiliation{$^1$ Department of Materials Science,
Raman Center for Atomic, Molecular and Optical Sciences,
Indian Association for the Cultivation of Science,
Jadavpur, Kolkata 700032, INDIA. \\
$^2$ Department of Physics, Oklahoma State University, Stillwater,  OK 74078, USA. }

\begin{abstract}
We show theoretically that it is possible to create and manipulate a pair of bound states in continuum in ultracold atoms by two lasers 
in the presence of a  magnetically tunable Feshbach resonance.  These bound states are formed due to coherent superposition of two electronically 
excited molecular bound states and a quasi-bound state in ground-state potential. 
These superposition states are decoupled from the continuum of two-atom collisional states. 
Hence, in the absence of other damping processes they are non-decaying. We analyze in detail the physical conditions that can 
lead to the formation of such states in cold collisions between atoms, and 
discuss the possible experimental signatures of such states.  
An extremely narrow and asymmetric shape with a distinct minimum of   
photoassociative absorption spectrum or scattering cross section as  a function of collision energy
will indicate the occurrence of a bound state in continuum (BIC). We prove that the minimum will occur at 
an energy at which the BIC is formed. 
We discuss how a BIC will be useful for efficient creation  
of Feshbach molecules and  manipulation of cold collisions.  Experimental realizations of BIC will pave the way 
for a new kind of bound-bound spectroscopy in ultracold atoms.  
\end{abstract}

\pacs{03.65.Ge,03.65.Nk,32.80.Qk,34.50.Rk}

\maketitle

\section{Introduction}

First introduced by von Neuman and Wigner more than eighty years ago  \cite{neuman1929}, 
a bound state in continuum (BIC) is a counter-intuitive and fundamentally profound concept. 
The original theoretical approach of Neuman and
Wigner  has undergone extensions and modifications over the years \cite{fonda1960,stillinger1975,pra32:1985:friedrich}.
In recent times, it has attracted  renewed research interests \cite{phystoday} with prospective applications
in many areas \cite{Capasso1992,prb:2006:bulgakov,prl:2009:moiseyev, 
prl107:2011:plotnik,nature499:2013:hsu}. A BIC refers to        
a discreet eigenstate with energy eigenvalue above the threshold of the continuum of a potential. 
The amplitude of the wave function of this state 
falls off in space and so  the wave function is  square-integrable. 
Normally, the eigenstates of a one-particle or a multi-particle 
system above the continuum  are infinitely  extended and   sinusoidal  at distances
larger than the range of the potential.  Below the threshold, 
there exists negative-energy spectrum of  discrete square-integrable bound states. 
The  idea of von Neuman and Wigner was to assume first the existence of a positive-energy square-integrable
wave function with its envelop decaying in space, and 
then to construct an appropriate potential that can support such states.
Physically, a BIC occurs due to destructive interference of the 
outgoing Schr\"{o}dinger waves scattered by the potential, creating an ``unusual'' trap  \cite{phystoday} for  
an electron   \cite{neuman1929}.  Hsu {\it et al.} \cite{nature499:2013:hsu} 
have observed trapped light,  namely, a BIC  of radiation modes by the destructive interference
of outgoing radiations amplitudes. 

Nearly forty five years after its discovery \cite{neuman1929},    
Stillinger and Herrick \cite{stillinger1975} extended the idea of  BIC to  two-body 
interactions, and discussed its applications in atomic and molecular physics. 
For two interacting particles, a BIC can be identified 
with a scattering resonance state with zero width. In general, a resonance at finite energies arises due to the existence of a 
quasi-bound (almost bound) state at positive energy. In the absence of any other source of dissipation, 
it is the coupling of the quasi-bound state  with the continuum of scattering states that results in 
finite width of the resonance. 
This means that  zero width of the resonance would imply decoupling of the quasi-bound state from the continuum 
of scattering states. In other 
words, the resonance state with zero width  becomes a BIC  \cite{fonda1960,pra32:1985:friedrich}.

Here we show that it is possible to create a BIC in cold atom-atom collisions in the presence of two 
photoassociation (PA) lasers near a 
magnetic field-induced Feshbach resonance. Our proposed scheme is depicted in Fig.1. The two lasers $L_1$ and $L_2$ are tuned near to 
the resonance of two excited molecular (bound) 
states $\mid b_1 \rangle $ and $\mid b_2 \rangle$, respectively. 
We consider a magnetic Feshbach resonance of two colliding ground-state atoms with two ground-state 
channels of which one is closed and the other open. In the absence of coupling with the open-channel, 
the closed channel is assumed to support a bound state $\mid b_c \rangle$. The two PA lasers couple  open-channel continuum 
 of scattering states $\mid E \rangle_{{\rm bare}} $ with  $E$ being collision energy, 
 and  $\mid b_c \rangle$ to both the excited bound states. 
Using projector operator techniques, we analyze the resolvent operator $(z - \hat{H})^{-1}$ of the Hamiltonian operator 
$\hat{H}$  and thereby arrive at an effective complex Hamiltonian $\hat{H}_{eff}$ of the three interacting 
bound states. $\hat{H}_{eff}$  is non-hermitian and its eigenvalues are in general complex. However, as we will 
demonstrate, under appropriate physical conditions, two of the eigenvalues of the effective Hamiltonian
can be made real. We establish the mathematical relations involving the parameters of our model that should hold 
good for the existence of the real eigenvalues. The
eigenvectors corresponding to the real eigenvalues  are non-decaying states and hence 
represent  bound states  in continuum. Similar effective Hamiltonians and their eigenvalue spectrum were 
studied in the context of two-photon dressed atomic continuum or autoionizing states \cite{pra33:1986:lami,jpb19:1986:kyrola,
pra:haan:1987} in 1980s.  In passing, 
we would like to mention that non-hermitian Hamiltonians with real eigenvalues also arise in other areas such 
parity-time (PT) symmetric Hamiltonian systems \cite{prl80:1998:bender,rpp70:2007:bender} 
and Friedrichs-Fano-Anderson model \cite{cpam1:1948:friedrichs,fano:1961,pr124:1961:anderson} 
where similar spectral singularity 
appears and can be associated with a non-decaying state in continuum \cite{prl102:2009:mostafazadeh,prb80:2009:longhi}.

Here we emphasize  that it is possible to detect the two predicted bound states in continuum by two 
spectroscopic methods, namely photoassociative absorption and photoassociative ionization techniques. Mathematically,
a BIC in our model appears as a spectral singularity in the scattering cross section as a function of energy. The expressions 
of photoassociation probability of either excited bound states as well as the scattering cross section 
involve the inverse  operator $(E - \hat{H}_{eff})^{-1}$. This means that for a real eigenvalue of $\hat{H}_{eff}$ 
the denominator  of the expressions goes to zero. This leads to divergence in scattering cross section implying the occurrence 
of a resonance with zero width \cite{fonda1960,pra32:1985:friedrich}, that is, a BIC.  However,  
photoassociative absorption spectrum does not diverge 
for a real eigenvalue, because the numerator of the expression for the spectrum also goes to zero for the real eigenvalue 
canceling out the singularity of  $(E - \hat{H}_{eff})^{-1}$. Physically, the singularity in scattering cross section 
implies that  BIC is a non-decaying state and hence decoupled from the continuum. However, BIC can make transition to either 
of the excited bound state via BIC-bound coupling leading to a finite probability for the absorption of a photon. Practically, 
spectral singularity can not be observed in an experiment. Instead, the signature of BIC will be manifested as an ultra narrow
in the coherent photoassociative spectrum or scattering cross section when the collision energy is tuned very close to the energy of the BIC. 
Usually, photoassociation is described in terms of atom loss from traps, 
due to the formation of excited diatomic molecules  decaying into two hot atoms or to a diatom  that can 
escape from trap.    
The occurrence of the bound states in continuum in cold collisions will facilitate to   
photoassociate two atoms effectively through a bound-bound transition process which can be coherent. We show that
a possible signature of a BIC in photoassociative cold collisions appears as a sharp  
and asymmetric line in photoassociative absorption spectrum as a function of collision energy. Close to the sharp spike-like 
line, there lies  a minimum which resembles to well-known Fano minimum \cite{fano:1961} and corresponds to 
the energy of the BIC.

We further demonstrate that, when the intensities and  the detuning  parameters of $L_1$ and $L_2$ are adjusted appropriately, 
one of the bound states in continuum can be reduced to a superposition of  $\mid b_1 \rangle $ and $\mid b_2 \rangle$ only while 
the other BIC results from superposition of all three bound states. We call the first one as A-type BIC and the second one as B-type BIC. 
The existence of A-type BIC can be probed by a probe laser producing the molecular ion and measuring the ion yield as a function of 
laser frequency. 
When the two continuum-bound couplings are much larger than the Feshbach 
resonance linewdith, the superposition coefficient of $\mid b_c \rangle $ state in B-type BIC is much larger than those of $\mid b_1 \rangle$ 
and $\mid b_2 \rangle$. Since the state $\mid b_c \rangle $ has a magnetic moment, B-type BIC can be probed by bound-free or 
bound-bound radio-frequency spectroscopy. In case of bound-free spectra,  
the final state would be two free atoms, and thus B-type BIC 
can be used for controlling  collisional properties of cold  atoms. Furthermore, 
Feshbach molecules can be created  by stimulating  
bound-bound transitions with a radio-frequency pulse at a fixed  magnetic field strength. To 
create Feshbach molecules, the usual method  uses a sudden sweep of magnetic field 
from large negative to large positive scattering length sides of  the  Feshbach resonance. 
However, this sudden sweep of magnetic field leads to substantial atom loss due to increase of 
kinetic energy and thereby limits the 
atom-molecule conversion efficiency.  In contrast, since a BIC is effectively decoupled from the continuum, 
by creating a B-type BIC, 
Feshbach molecules can be produced efficiently by inducing stimulated transitions from the BIC 
to Feshbach molecular states with a radio-frequency pulse.

We also show that, when the coupling of the continuum to one 
the bound states is turned  off, one can still find one BIC  which can be identified as 
the familiar ``dark state'' made of 
superposition of the two remaining bound states.  Coherent population trapping occurs in this superposition 
state 
resulting in the vanishing of the probability of the continuum of scattering states. 
When laser coupling to either of the 
excited states is turned off, the model reduces to one \cite{jpb:2009:deb} 
that describes Feshbach resonance-induced 
Fano effect in photoassociation. The effective Hamiltonian approach to this model
shows that the BIC appears  at an energy at which Fano minimum occurs.  This can be identified 
with the standard result that the population trapping occurs due to the 
``confluence`` of coherences \cite{prl47:1981:eberly} at Fano minimum. 
 When the quasibound state in ground-state potential 
is absent or the magnetic Feshbach resonance is turned off, the resulting 
effective Hamiltonian has a real eigenvalue when the corresponding eigenvector 
is an excited molecular dark state \cite{pra:2014:subrata}.

The paper is organized in the following way. In Secs. II and III, we present our model and its solution, respectively. 
We analyze in some detail how to realize our model and its application in cold atoms and molecules in Sec. IV. Finally, 
we discuss important conclusions of our study in Sec. V.

\section{The model}

The model is schematically depicted in Fig.1. To begin with, we keep our model most general. 
Suppose a two-channel model  is capable of 
describing an $s$-wave  Feshbach resonance in ground-state atom-atom cold collision. One of these two channels is open and the other 
is closed.  The closed channel is assumed to support a  bound state $\mid b_c \rangle $. 
The  thresholds of these two ground-state channels and the binding energy of $\mid b_c \rangle $
are tunable with an external magnetic field. 
Both the bare continuum of scattering states $\mid E \rangle_{{\rm bare}}$ in the open channel with $E$ being the collision energy, 
and $\mid b_c \rangle$ are coupled to two  bound states $\mid b_1 \rangle$ and $\mid b_2 \rangle$ in an excited molecular 
potential by two lasers 
$L_1$ and $L_2$, respectively.
Suppose, the states  $\mid b_1 \rangle$ and $\mid b_2 \rangle$ have 
same rotational quantum numbers $J_1 = J_2 = 1$, but they have different vibrational quantum numbers if both of them are supported 
by the same adiabatic molecular potential.  In case they belong to different molecular potentials, their vibrational quantum 
numbers may be same or different. The energy spacing between $\mid b_1 \rangle$ and $\mid b_2 \rangle$ is assumed to be large enough compared to the line widths of 
the two lasers. Furthermore, $\mid b_1 \rangle$ and $\mid b_2 \rangle$ are assumed to be far below the  dissociation threshold 
 of excited potential(s) so that the transition
 probability at the single-atom level be negligible. 
 
 \begin{figure}
\includegraphics[width=3.5in]{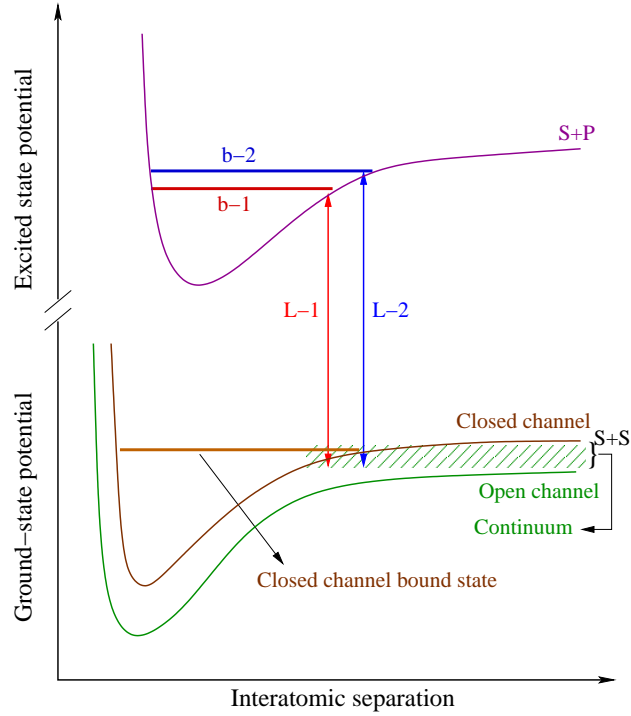}
\caption{A schematic diagram for creating BIC in ultracold atoms. Two lasers L$_1$ and L$_2$ are used to excite PA transitions
from the magnetic Feshbach-resonant collisional state of two ultracold ground-state ($S + S$) atoms to the 
two bound states $b_1$ and $b_2$, respectively,  in the same electronically 
excited molecular potential. The magnetic Feshbach resonance is considered as a two-channel model in the electronic
ground-state potentials with the lower channel being open and the upper one being closed. In the large separation limit 
the ground-state channel potential corresponds to two separated  $S + S$ atoms while the excited-state potential connects 
to two separated $S + P$ atoms. } 
\end{figure}
 
In the rotating wave approximation, the Hamiltonian of our mode  can be expressed as
$\hat{H} = \hat{H}_0 + \hat{V}$ where 
\begin{widetext}
\bea
 \hat{H}_0 &=&  \sum_{n} (E_{n} - \hbar \omega_{L_n}) \mid b_n \rangle \langle b_n \mid 
+ E_{0} \mid b_c \rangle \langle b_c \mid
+  \int E' dE' \mid E' \rangle_{\rm{bare}} \, _{\rm{bare}} \langle E' \mid 
\eea
\bea 
\hat{V} =  \sum_{n}  \int d E' \Lambda_{n}(E')  
\mid b_n \rangle \, _{\rm{bare}}\langle E'\mid  + 
\int d E' V_{E'} \mid b_c \rangle \, _{\rm{bare}}\langle E'  \mid 
+ \sum_{n} \hbar \Omega_{n} \mid b_n \rangle \langle b_c \mid + \rm{C.c.}  
\eea 
\end{widetext}
 $E_n$ is 
the binding energy of $n$-th excited molecular state $\mid b_n \rangle$, $\omega_{L_n}$ denotes 
the frequency of the $L_n$ laser, 
$E_{c}$ being the energy of the cosed channel bound state $\mid b_c \rangle$ and 
$\mid E'\rangle_{\rm{bare}}$ the  bare continuum of scattering state with energy  $E'$. Note that all the energies are measured from the 
open channel threshold unless stated otherwise.
Here $ \Lambda_{n}(E)$ is the dipole
matrix element of  transition $  \mid E \rangle_{\rm{bare}} \rightarrow  \mid b_n \rangle $, 
$V_{E}$ is the coupling between the closed channel bound state $\mid b_c \rangle $  and the open channel scattering state 
$\mid E \rangle_{{\rm bare}}$, 
and $\Omega_{n}$ is the Rabi frequency between  $\mid b_n \rangle $ and $\mid b_c \rangle$. 
  The magnetic Feshbach resonance  linewidth is  
 $\Gamma_f = 2 \pi |V_E|^2$. 

To study BIC, we analyze the resolvent operator $G(z) = (z- \hat{H})^{-1}$ and introduce the projection operators 
\bea 
P = \mid b_c \rangle \langle b_c \mid + \sum_{n=1,2} \mid b_n \rangle \langle b_n \mid
\eea 
\bea Q = 1 - P = \int d E \mid E\rangle_{\rm{bare}} \, _{\rm{bare}} \langle E \mid \eea
which satisfy the properties 
\bea 
P P = P, \hspace{0.2cm} Q Q = Q, \hspace{0.2cm} P Q = Q P = 0, \hspace{0.2cm} P + Q = 1
\eea 
Thus we have 
\bea 
G = G_0 + G_0 \hat{V} G  =  \frac{1}{E - \hat{H}_0 + i\epsilon} + \frac{1}{E - \hat{H}_0 + i\epsilon} \hat{V} G
\eea 
Projecting out the bare continuum states, after some algebra as given in appendix-A, we obtain an effective
 Hamiltonian of interacting three bound states. Explicitly, this Hamiltonian is given by  
\begin{widetext}
\bea 
H_{eff} &=& H_0 + 
\sum_{n,n' = 1,2} \left [  \left ( \hbar \delta_{n n'} - i \frac{\hbar \Gamma_{n n'}(E)}{2} \right )
\mid b_n \rangle \langle b_{n'} \mid + 
\left (\delta_c  - i \frac{\hbar \Gamma_{f}(E)}{2} \right ) \mid b_c \rangle \langle b_c \mid 
\right. \nonumber \\
&+&  \left. \sum_{n}  \frac{\hbar \Gamma_{n f}(E)}{2} \left \{ q_{nf} - i  \right \}
 \mid n \rangle \langle b_c \mid +
 {\rm{C.c.}} \right ] \nonumber \\
\eea 
\end{widetext}
where   
$ \delta_{n n}  = (E_n + \Delta_{n n}^{\rm{shift}})/\hbar - \omega_{L_n}$ is the detuning of the light-shifted 
$n$th excited level from the $L_n$ laser frequency $\omega_{L_n}$, 
$\delta_{n n'} =  \hbar^{-1} \Delta_{n n'}^{\rm{shift}}$ ($n \ne n'$), where $\Delta_{n n'}^{\rm{shift}}$
is the real part of the quantity $ \int d E' \Lambda_n^*(E') \Lambda_{n'}(E')/(E - E')$ 
between, $\Gamma_{n n'}(E) = 2 \pi \Lambda_n^*(E) \Lambda_{n'}(E)$. Here  
${\delta}_c = \hbar^{-1} \left [E_{c}(B) +  \Delta_{f}^{\rm{shift}} - E_{th}(B) \right ] r$ is the detuning of the 
shifted closed channel bound state level from the the threshold $E_{th}$ of the open channel.
Note that $\delta_c$ is a function of the applied magnetic field $B$ due to the dependence of $E_c$ 
and $E_{th}$ on $B$.
$\Delta_{f}^{\rm{shift}} = {\mathcal P} \int d E' |V_{E'}|^2/(E - E')$, where  ${\mathcal P}$ 
stands for Cauchy's principal value, is the shift due to magnetic coupling 
$V_{E'}$ between $\mid b_c \rangle$ and $\mid E' \rangle_{{\rm bare}}$. $q_{n f}$ is the well-known Fano-Feshbach 
asymmetry parameter defined by 
\bea 
q_{n f} = \frac{\delta_{nf}^{\rm{shift}} + \Omega_n}{\Gamma_{nf}/2}.
\eea
where $\Omega_n$ is the Rabi frequency for transition 
$\mid b_c \rangle \leftrightarrow \mid b_n \rangle$
and 
\bea 
\delta_{nf}^{\rm{shift}} = \hbar^{-1} {\mathcal P} \int d E' \Lambda_{n}^*(E') V_{E'}/(E - E')
\eea
is a frequency-shift of  $\mid b_c \rangle \leftrightarrow \mid b_n \rangle$ transition frequency 
due to the indirect coupling of the two bound states via the continuum. 
Here $\Gamma_{n f} = 2 \pi \hbar^{-1} \Lambda_{n}^*(E) V_{E}$. 
For energy $E$ near the threshold of the continuum, 
the region $E' > E $ of the above integrand contributes  more strongly 
\cite{pra:1999:bohn}. As a result,  $\delta_{nf}^{\rm{shift}}$ 
will be negative at low energy. Since $\Omega_n$ is positive, the sign of $q_{nf}$ depends on the relative 
strength between $\left | \delta_{nf}^{\rm{shift}} \right |$ and $\Omega_n$. Since the magnetic coupling $V_{E'}$ is determined 
by the hyperfine spin coupling between the closed and the open channel, its value depends on the specific 
atomic system chosen. In contrast, the laser couplings $\Omega_n$ and $\delta_{nf}^{\rm{shift}}$ depend 
 on which bound state $ \mid b_n \rangle$ is chosen for the laser to be tuned to, in accordance with Franck-Condon 
 principle of molecular spectroscopy. Thus, it is possible to alter the sign and magnitude of Fano-Feshbach asymmetry 
 parameter in our model through the selectivity of the excited molecular bound states. Since in molecular excited 
 states, there is a host of vibrational levels in different molecular symmetries that can be accessed by PA, there is a 
 lot of flexibility in choosing the excited bound states in our model. We will further discuss this point 
 in Sec.IV.

In the absence of the 
lasers fields, the magnetic field-dependent resonant scattering phase shift $\eta_r$ and the $s$-wave scattering 
length $a_s$ are given  by 
\bea 
- \cot \eta_r = \frac{E - \tilde{E}_c}{\hbar \Gamma_f/2} \simeq \frac{1}{k a_s} +   \frac{1}{2} r_0 k 
\eea 
$\tilde{E} = E_{c} +  \Delta_{f}^{\rm{shift}}$ is the shifted energy of $\mid b_c \rangle$, $k$ is the wave number
related to the collision energy $E = \hbar^2 k^2/2\mu$ with $\mu$ being the reduced mass of the two colliding atoms. Here 
$r_0$ is the effective range of the open-channel ground-state potential. In the limit $k \rightarrow 0$, 
$  \Gamma_f/2 \simeq k G_f $ where $G_f$ is a constant having the dimension L s$^{-1}$. The scattering length $a_s$ 
and the effective range $r_0$ are related to $\tilde{E}_c$ and $G_f$ by 
$ 
\frac{1}{a_s} = \frac{-\tilde{E}_c}{\hbar G_f}$ and  
$
r_0 = \frac{\hbar}{\mu G_f} 
$, respectively.  This means the magnetic field dependent detuning $\delta_c(B) = - G_f/a_s$. When $\tilde{E}_c > 0 $, $a_s$ 
is negative and the $\mid b_c \rangle$ lies above the threshold of the open channel and hence $\mid b_c \rangle$ 
is a quasi-bound state. In contrast, when $\tilde{E}_c < 0$, $\mid b_c \rangle$ is a true bound state (Feshbach molecular 
state) and scattering length is positive. Later, we will show that by forming a BIC with large scattering length, the BIC 
can be converted into a Feshbach molecule by stimulated radio-frequency spectroscopy.

\section{The solution} 

For simplicity, let us  introduce the dimensionless parameters $\tilde{\delta}_{n} = \delta_{n n}/(\Gamma_f/2)$ 
 $g_{n}= \Gamma_{nn}/\Gamma_f$, $g_{n n'} = \Gamma_{n n'}/(\Gamma_f/2)$  ($n \ne n'$).  
We assume that 
$\delta_{1 2}^{\rm{shift}} = \delta_{ 2 1}^{\rm{shift}} \simeq 0$, that is, 
the real parts of laser-induced couplings 
between $\mid b_1 \rangle$ and $\mid b_2 \rangle$  are  negligible. Assuming 
the two free-bound photoassociative couplings $\Lambda_{n E}$ to be real quantities, 
we have $ \Lambda_{n E} V_E/|V_E|^2 = \Lambda_{n E} /V_E = \sqrt{g_n}$
Under these conditions, the effective Hamiltonian of Eq.(7)  can be written in matrix form
\bea 
H_{eff} =  \frac{\hbar \Gamma_f}{2} \left [ \mathbf{A} + i \mathbf{B} \right ]
\eea 
where 
\be 
\mathbf{A} = \left(
\begin{array}{c c c}
\tilde{\delta}_1 & 0  & q_{1 f} \sqrt{g_1} \\
 0 & \tilde{\delta}_2 & q_{2 f} \sqrt{g_2} \\
q_{1 f} \sqrt{g_1} & q_{2 f} \sqrt{g_{2}} & - (k a_s)^{-1}  
\end{array}
\right ) 
\ee 
and 
\be 
\mathbf{B} =  \left(
\begin{array}{c c c}
- g_1 & - g_{12}  & - \sqrt{g_{1}} \\
- g_{21} & - g_2 & - \sqrt{g_{2}} \\
- \sqrt{g_{1}} & - \sqrt{g_{2}} & - 1 
\end{array}
\right ) 
\ee
For $(k a_{s})^{-1} = 0$,  these matrices have the same form as
the Eq. (2.18) of Ref.\cite{pra:haan:1987}. The secular equation for ${\mathbf B}$ matrix 
is $ 
x^3 + (g_1 + g_2 + 1 ) x^2  = 0 $ and it
has two roots equal to zero and the third one equal to $-(g_1 + g_2 + 1)$.
When the two eigenvectors of ${\mathbf B}$ with zero eigenvalues become simultaneous 
eigenvectors of ${\mathbf A}$ with real eigenvalues, we  have 
two real roots of the effective Hamiltonian. In addition,  when  ${\mathbf A}$ and 
${\mathbf B}$ commute, both these matrices are diagonalizable within simultaneous eigenspace, with 
$H_{eff}$ having two real eigenvalues. The commutative condition can be easily found to be 
\bea 
q_{1 f} + \tilde{\delta}_1 = q_{2 f} + \tilde{\delta}_2 = q_{1 f} g_1 + q_{2 f} g_2 - (k a_s)^{-1}
\label{biccond}
\eea

To evaluate the two real eigenvalues of the complex Hamiltonian $H_{eff}$, we proceed in the following way. 
We first get an eigenvector of the matrix ${\mathbf A}$ with an unknown eigenvalue $\lambda$ in the form 
\bea 
X = C \left (  \begin{array}{c} 
 1 \\
x_2 \\   
 x_3  \end{array} \right )
\eea 
where $C$ is   normalization constant and $x_2$ and $x_3$ are the two elements of the vector. All 
these three quantities $C$, $x_2$ and $x_3$ are the functions of $\lambda$. 
Assuming that $X$ is also an eigenvector of $B$ with zero eigenvalue, 
the  eigenvalue equation ${\mathbf B} X = 0 $ leads to a quadratic equation 
for $\lambda$, the solutions of which are the desired eigenvalues.  
For $(k a_s)^{-1} = 0$, that is, for $a_s \rightarrow \infty$ or for the magnetic field tuned 
on the Feshbach resonance, the two real eigenvalues of $H_{eff}$ are $E_{\pm} = \lambda_{\pm} \hbar \Gamma_f/2 $ 
where
\bea 
\lambda_{\pm} = \frac{1}{2}(\tilde{\delta}_2 - q_{1f}) \pm  \frac{1}{2}
\left [ \left (\tilde{\delta}_2 + q_{1f} \right )^2 - 4 g_2 q_{2 f} (q_{1f} - q_{2f}) \right ]^{1/2}
\eea 
provided $ \left ( \tilde{\delta}_2 + q_{1f} \right )^2 \ge 4 g_2 q_{2 f} (q_{1f} - q_{2f})$. $\lambda_+$ 
and $\lambda_-$ are the two eigenvalues of ${\mathbf A}$, the corresponding eigenvectors are also the 
eigenvectors of ${\mathbf B}$ with both eigenvalues being zeros. 
The two eigenstates corresponding to these two real eigenvalues are the coherent superpositions of 
the three bound states, and represent two bound states in continuum. Note that the two real eigenvalues are expressed 
in terms of $g_2$, $\tilde{\delta}_2$ and the two Fano-Feshbach asymmetry parameters. But, both the remaining 
parameters $\tilde{\delta}_1$ and $g_1$ can not  be arbitrary when  $H_{eff}$  has the real eigenvalues. By expressing 
$x_2$, $x_3$ and $C$ in terms of the set of the parameters  $g_2$, $\tilde{\delta}_2$, $q_{1 f}$ and $q_{2 f}$, from the 
eigenvalue equation ${\mathbf A} X = \lambda X$ one obtains 
\bea 
  g_1 
  =  \frac{ \left ( \tilde{\delta}_1 - \lambda  \right ) \left (  \tilde{\delta}_2 - \lambda - g_2 q_{2 f}  \right ) } 
  { q_{1f} \left ( \tilde{\delta}_2 -  \lambda  \right)  } 
\eea 
with $\lambda \ne \tilde{\delta}_2$. 
Now, replacing $\lambda$ by a real eigenvalue of Eq. (16), one 
can use the above equation to set the appropriate parameter space of $g_1$ and  
 $\tilde{\delta}_1$ for which  $\lambda$  remains real and fixed for a fixed set of other parameters. 
  
Let us now consider the special case of  both excited bound states belonging to the same excited molecular potential
with  closely lying vibrational quantum numbers  
$1 \le |v_1 - v_2 | \le 2$. Hence,  the bound-bound Franck-Condon (FC) factors  
for transitions $\mid b_c \rangle \leftrightarrow \mid b_1 \rangle$ and $\mid b_c \rangle \leftrightarrow \mid b_2 \rangle$ 
will be nearly equal. Similarly, the  free-bound FC factors for 
transitions $\mid E \rangle_{{\rm bare}} \leftrightarrow \mid b_1 \rangle$ and 
$\mid E \rangle_{{\rm bare}} \leftrightarrow \mid b_2 \rangle$ will also be almost equal. 
Since Fano asymmetry parameters $q_{1 f}$ and $q_{2 f}$ are independent of laser intensities, but 
are dependent of these  FC factors, we expect $q_{1 f} \simeq q_{2 f} $.  
 The BIC condition of Eq.(\ref{biccond}) then implies 
$\tilde{\delta}_1 = \tilde{\delta}_2$. 
Now, putting $q_{1 f} = q_{2 f} = q_f$, the commutativity condition implies $\tilde{\delta}_1 = \tilde{\delta}_2  = q_f (g_1 + g_2 -1 ) $.
 Under these conditions, two real roots of the effective Hamiltonian are 
\bea 
E_{A} = \frac{\hbar \Gamma_f}{2} \lambda_{+} = \frac{\hbar \Gamma_f}{2} q_f (g_1 + g_2 -1 )  \\
E_{B} = \frac{\hbar \Gamma_f}{2} \lambda_{-} = -  \frac{\hbar \Gamma_f}{2} q_f 
\eea 
The BIC state corresponding to $E_A$ is 
\bea 
\mid A \rangle_{{\rm BIC}} = \frac{1}{\sqrt{g_1 + g_2}} \left [ \sqrt{g_2} \mid b_1 \rangle - \sqrt{g_1} \mid b_2 \rangle \right ]
\eea 
Note that this state  does not mix with  the closed channel 
bound state $\mid b_c \rangle$, and so this eigenvector  
is immune to magnetic field tuning of the Feshbach resonance.  Nevertheless,  $\mid b_1 \rangle $ and $\mid b_2 \rangle$ remain coupled 
with $\mid b_c \rangle$ and $\mid E \rangle_{{\rm bare}}$ due to the lasers.
It is easy to see that the photoassociative transition matrix element for the interaction Hamiltonian 
$\hat{V}_{PA} = \sqrt{g_1} \mid b_{1}\rangle \, _{bare} \langle E \mid + \sqrt{g_2} \mid b_{2}\rangle \, _{bare} \langle E \mid +{\rm C.c.}$ between 
 $\mid E \rangle_{{\rm bare}}$ and $\mid A \rangle_{{\rm BIC}}$ is zero. This means that this is an excited molecular 
dark state  that is predicted to play 
an important role in suppression of   photoassociative atom loss \cite{pra:2014:subrata}. 
We call this dark  state as A-type BIC. 
The eigenstate corresponding to the eigenvalue $E_B$  is given by 
\bea 
\mid B \rangle_{{\rm BIC}} = \left [\frac{1}{(g_1+g_2)(g_1+g_2+1)}\right ]^{\frac{1}{2}} \left [ \sqrt{g_1} \mid b_1 \rangle + \sqrt{g_2} 
\mid b_2 \rangle + (g_1 + g_2) \mid b_c \rangle \right ] 
\label{b-type} 
\eea 
This is a  superposition of all three bound states. The involvement  of $\mid b_c \rangle $ makes this BIC dependent  
on the magnetic field $B$. We call this state as B-type BIC. Near 
unitarity regime,  $|k a_s|$ is large and consequently $| (k a_s)^{-1}| <\!< 1$, and hence the effect of finite $(k a_s)^{-1}$ can be taken into account 
perturbatively. The perturbation part of the Hamiltonian is then $V = - \frac{\hbar \Gamma_f }{2} (k a_s)^{-1} \mid b_c \rangle \langle b_c \mid $, and the first order 
correction to the energy $E_{B}$ is given by 
\bea 
\Delta E_{B} = \, _{{\rm BIC }} \langle B \mid V \mid B \rangle_{{\rm BIC }} =  - \frac{\hbar \Gamma_f }{2} (k a_s)^{-1} \frac{g_1 + g_2}
{g_1 + g_2 + 1} 
\eea 

The signature of this BIC can be detected in a number of coherent spectroscopic methods as will be  discussed
in the next section. For example, a BIC  may be manifested as a strong and narrow photoassociative absorption line.

\section{Applications: Results and discussions}

Before we discuss some specific applications, it is worthwhile to make some general observations on the 
dependence of the two real eigenvalues $E_A$ and $E_B$ on the magnetic field tuning of Feshbach resonance. In the 
zero energy limit ($E \rightarrow 0$) and near the vicinity of Feshbach resonance, 
the applied magnetic field $B$ and the scattering length $a_s$ are related by 
\bea 
a_s^{-1} = - a_{bg}^{-1} \frac{B - B_0}{\Delta} 
\eea 
where $B_0$ is the resonance magnetic field at which $a_{s} \rightarrow \infty$ and $a_{bg}$ is the background scattering 
length. Since $a_{bg} \Delta >0$,  $a_s < 0$ for $B > B_0$ and $a_s > 0$ for $B < B_0$. In case of Fermionic atoms, 
$B > B_0$ ($B < B_0$) region is commonly known as BCS (BEC) side of the resonance. The parameter range
$-1.0 \le (k a_s)^{-1} \le 1.0$ is usually referred to as `unitarity' regime. 

Though $E_A$ changes with the change of $(k a_s)^{-1}$, the corresponding eigenstate of Eq. (20) 
remains intact and so  coherent population trapping occurs (CPT) in A-type BIC that remains protected against the tuning 
of magnetic field or the scattering length. In contrast, both eigenvalue and eigenstate of B-type BIC depends on
$B$ or $a_s$. For $a_s \rightarrow \infty$ and $g_1 + g_2 > 1$, as $q_f \rightarrow \pm 0 $, 
$E_A \rightarrow \pm 0$ and $E_B \rightarrow \mp 0$ as can be inferred from the expressions (18) and (19).  Note that 
the eigenvalues of both A- and B-type BIC depend inversely on $k a_s$.

\subsection{Detection of BIC via photoassociation} 

Modifications of photoassociation probability  as a result of the formation of BIC  
can be ascertained by making use  of isometric 
and invertible M\o ller operators $\Omega_{\pm}$ of scattering theory.  Since before turning on the 
lasers and the magnetic field, the atoms are in a resonant collisional state, 
incoming  state of the problem can be taken
to be the bare continuum 
$\mid E \rangle_{{\rm bare}} $.  The dressed  continuum state $\mid E+ \rangle$  is given by 
\bea 
\mid E+ \rangle = \Omega_{+} \mid E \rangle_{{\rm bare}} 
\eea 
where 
\bea 
\Omega_{+} = 1 + G(z + i \epsilon ) V 
\eea 
The  probability of photoassociative transition $\mid E \rangle \rightarrow \mid b_n \rangle $ is given by 
\bea 
P_n = \int d E |\langle b_n \mid E+ \rangle |^2
\eea  
The quantity 
\bea 
 S_n(E) = |\langle b_n \mid E+ \rangle |^2
 \eea 
 is the photoassociation probability per unit collision energy. Now, we have 
\bea 
\langle b_n \mid  E + \rangle &=& \langle b_n \mid (P+Q) G(z + i\epsilon)  (P+Q) V \mid E \rangle \nonumber 
\\
&=& \langle b_n \mid P G(z + i\epsilon)  (P+Q) V \mid E \rangle_{{\rm bare}}
\eea
Using Eq. (A.3) we have 
\bea 
P G Q &=&  P  \left ( Q + G P V Q \right ) \frac{1}{E - H_0 - V Q + i\epsilon} \nonumber \\
&=& PGP V Q \frac{1}{E - H_0 - V Q + i\epsilon} 
\eea 
\bea
\langle b_n \mid \Omega_{+} \mid E \rangle_{{\rm bare}} &=&
\langle b_n \mid  PG(E + i \epsilon) P R(E + i \epsilon)  \mid E \rangle_{{\rm bare}} 
\eea 
where $R(E + i\epsilon)$ is given in Eq. (A5). 
Now, \bea  PG(E + i \epsilon) P = (E - H_{eff} + i \epsilon )^{-1} =  \frac{2}{\hbar \Gamma_f} 
\frac{{\mathscr A}}{{\rm Det}[(\tilde{E} - \tilde{H}_{eff}]} \eea  
where $\tilde{E} = 2 E/\hbar \Gamma_f $, $\tilde{H}_{eff} = \mathbf{A} + i \mathbf{B}$ and 
${\mathscr A}$ is the transpose of the co-factor matrix of $(\tilde{E} - \tilde{H}_{eff} )$. Thus we have 
\bea 
\langle b_n \mid \Omega_{+} \mid E \rangle = \sqrt{\frac{2}{\pi \hbar \Gamma_f}} 
\frac{1}{{\rm Det}[(\tilde{E} - \tilde{H}_{eff}]}
\left [
\sum_{m=1,2} {\mathscr A}_{n m }
  \sqrt{g_m} + {\mathscr A}_{n 3} \right ]   
\eea
The quantity within the third bracket in the numerator of the above equation, for $n=1$ and  $(k a_s)^{-1}=0$,  can be expressed as 
\bea 
&-& \sqrt{g_1} \left [ - \tilde{E} (\tilde{\delta}_2 - \tilde{E})  - g_2 q_{2f}^2  - i (\tilde{\delta}_2 - \tilde{E})
+ i g_2 q_{2 f} 
+ i g_2 q_{2 f}  \right ] \nonumber \\
&+& \sqrt{g_2} \left [ - q_{1 f} q_{2f} \sqrt{g_1 g_2}  + i \sqrt{g_1 g_2} (q_{1 f} + q_{2 f}) + i \sqrt{g_1 g_2} \tilde{E} 
\right ] \nonumber \\
&-& \left [ - q_{1f} \sqrt{g_1} (\tilde{\delta}_2 - \tilde{E}) - i g_2 \sqrt{g_1} q_{2f} + i g_2 \sqrt{g_1} q_{1 f} +
i \sqrt{g_1} (\tilde{\delta}_2 - \tilde{E})
\right ] 
\eea 
the zeros of which are the roots of the quadratic equation  
\bea 
  \tilde{E} (\tilde{\delta}_2 - \tilde{E})  +  g_2 q_{2f}^2   - q_{1 f} q_{2f}  g_2   
+ q_{1f}  (\tilde{\delta}_2 - \tilde{E})   = 0 
\eea 
 Thus, the numerator has two zeros at 
\bea 
\tilde{E}_{\pm} =  \tilde{\delta}_2  +  \frac{1}{2} \left [ - (\tilde{\delta}_2 + q_{1 f})  \pm \sqrt{  (\tilde{\delta}_2 + q_{1 f})^2 
+ 4 q_{2 f} g_2 (q_{ 2f} 
- q_{1 f} )} \right ]  
\eea 
It is important to note that the two zeros of the numerator are the same as the two real eigenvalues of $H_{eff}$ as given in 
Eq. (16). This means that,  although the denominator 
\bea 
{\rm Det} \left [\tilde{E} - \tilde{H}_{eff} \right ] = \sum_{i=1}^{3} \left ( \tilde{E} - \tilde{E}_i \right ) 
\eea 
where $\tilde{E}_i$ represents an eigenvalue of $\tilde{H}_{eff}$, 
may become zero for a real eigenvalue of $\tilde{H}_{eff}$, the spectrum 
remains finite in the limit $\tilde{E} \rightarrow E_{i}$ for a real eigenvalue 
$\tilde{E}_i \equiv \lambda_{\pm}$. When 
the real part of a complex eigenvalue is nearly equal to a real eigenvalue, and the imaginary part 
 is extremely small ($<\!< \hbar \Gamma_f$), 
the spectrum $S(E)$ as a function of collision energy $E$ will exhibit   
Fano-like minimum  and  a highly prominent maximum lying close to the minimum. 
Experimentally, searching for such a spectral structure is possible by 
choosing the parameters $\tilde{\delta}_2$, $g_2$, $q_{1 f}$ and $q_{2 f}$ for a zero of the numerator, but $g_1$ and 
$\tilde{\delta}_1$ are chosen such that the real part of one of the eigenvalues of $H_{eff}$ is nearly equal to a  
zero of the numerator with the imaginary part being  small.  A Fano-like 
spectral structure with a nearby narrow spectral spike will indicate the existence of BIC \cite{pra:haan:1987}.  

\begin{figure}
\includegraphics[width=3.5in]{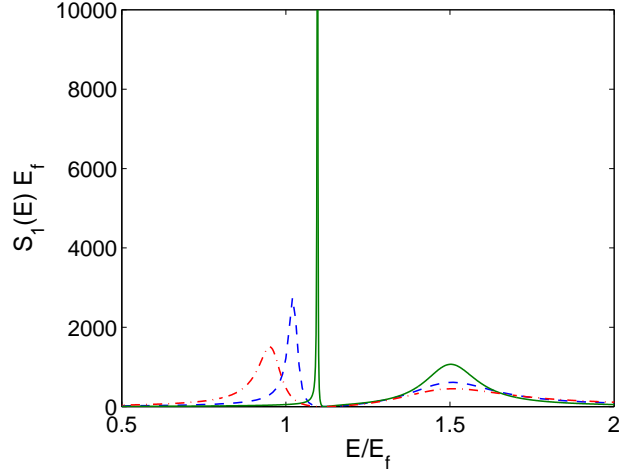}
\caption{ Dimensionless spectrum $S_1 (E) E_f $ as a function of dimensionless energy $E/E_f$ 
(where $E_f = \hbar \Gamma_f/2$)
for  $g_1 = 0.25 $ (solid), $g_1 = 0.5$ (dashed) and $g_1 = 0.75 $ (dashed-dotted)
with  $q_{1 f} = - 0.5$, $q_{2 f} = -1.0$, $g_2 = 2.0$, $\tilde{\delta}_2 = -0.5$ and 
$\tilde{\delta}_{1} = 1.5$. For the solid curve the three complex eigenvalues of 
$\tilde{H}_{eff}$ are $\tilde{E}_{1} = 1.0964 - 0.0010 i$, 
$\tilde{E}_2 = 1.4989 - 0.0992 i $, and $\tilde{E}_3 = -1.5953 - 3.1498 i$; for dashed curve 
$\tilde{E}_{1} = 1.0242 - 0.0164 i $, 
$\tilde{E}_2 = 1.4862 - 0.1653 i $, and $ \tilde{E}_3 = -1.5104 - 3.3184 i $; for dashed-dotted curve 
$ \tilde{E}_{1} = 0.9581 - 0.0421 i $, 
$\tilde{E}_2 = 1.4645 - 0.2111 i $, and $\tilde{E}_3 = -1.4226 - 3.4968 i $. When $g_1 = 0.1803$, $\tilde{E}_1 $
becomes real and is equal to 1.1180 which correspond to the minimum of the spectral lines. The spike height of 
the solid curve is $2.2 \times 10^4$. 
} 
\end{figure}

\begin{figure}
\includegraphics[width=3.5in]{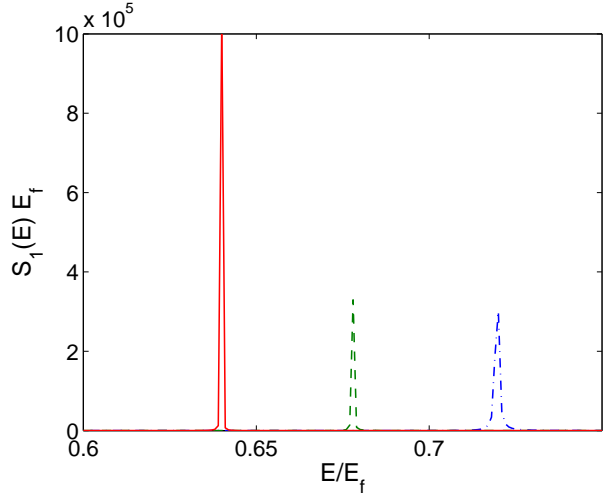}
\caption{ Same as in Fig.2 but  for $ g_1 = 5.5$ (solid), $g_1 = 5.0 $ (dashed) and $g_1 = 4.5 $ (dashed-dotted)
 with fixed parameters   $q_{1 f} = 0.5$, $q_{2 f} = 1.0$, $g_2 = 5.0$, 
 $\tilde{\delta}_1 = 1.45$ and $\tilde{\delta}_2 = - 1.5$. For solid curve the eigenvalues 
 are $\tilde{E}_{1} =  0.6645 - 4.5 \times 10^{-6} i$, 
$\tilde{E}_2 = -2.2178 - 0.4279 i$, and $\tilde{E}_3 =1.55338 - 11.0720 i$; 
 for dashed curve 
 $\tilde{E}_{1} =  0.7036 - 0.0002 i$, 
$\tilde{E}_2 = -2.2165 - 0.44027 i$, and $\tilde{E}_3 = 1.51292 - 10.5595 i$; for dashed-dotted 
curve $\tilde{E}_{1} =  0.7466 - 0.0008 i$, 
$\tilde{E}_2 = -2.2153 - 0.4534  i$, and $\tilde{E}_3 =1.4687 - 10.0457 i$. For $g_1 = 5.2516$, $\tilde{E}_1$ is real and 
equal to 0.6583. The spike height of the solid curve is $1.0 \times 10^6$. } 
\end{figure}

Figures 2 and 3 display  photoassociative absorption spectra $S_1(E)$ (which is a measure 
of probability of transition to $\mid b_1 \rangle$)  as a function function of  $E$ 
for different values of $g_1$ keeping all other parameters fixed.
We scale all energy quantities by $E_f = \hbar \Gamma_f/2$. 
The fixed parameters are chosen such that they fulfill 
the Eq. (35). The different values of $g_1$ are chosen close to a value that is given by the condition (17) for the occurrence 
of a real eigenvalue $\lambda = \tilde{E}_{\pm}$ for the given $\tilde{\delta}_1$. 
The results displayed in Figs. 2 and 3 
and the data mentioned in the figure captions clearly support our analytical results described above. 
The spikes in solid 
curves occur due to the BIC with real eigenvalue close to the real part of  $\tilde{E}_1$ . 
In Fig. 2, the bumps near $E = 1.5 E_f$ occurs due to the complex 
eigenvalue $\tilde{E}_2$. Note that, as can be seen from the Eq.(8), 
the Fano-Feshbach asymmetry parameter will be positive (negative) when the Rabi frequency $\Omega$ is greater (smaller)
than 
the magnitude $\delta_{n f}^{{\rm shift}}$  which is negative. 

\subsection{Possible realizations of the model }

We next discuss the possibility of experimental realization of our model in ultracold atomic gases that 
are of current experimental interests. The discussed BIC  can be realized in ultracold atoms with 
currently available experimental techniques of magnetic Feshabch resonances \cite{rmp:2010:grimm}
and photoassociation \cite{rmp:1999:weiner,rmp:2006:jones}. 
In particular, 
our theoretical proposal can be  implemented 
in case of  experimentally observed 
narrow Feshbach resonances in cold alkali atoms like $^{23}$Na
\cite{nature:1998:inouye,prl:1999:stenger, 
prl83:1999:verhaar,
pra:2000:tiemann, prl:2003:xu, prl:2004:mukaiyama}, $^{87}$Rb \cite{prl89:2002:marte,pra68:2003:volz,
pra70:2004:durr}, 
$^6$Li \cite{prl91:2003:hulet,prl91:2003:zwierlein}, $^7$Li \cite{nature:2002:strecker,prl102:2009:hulet},  etc..  Narrow or close-channel dominated Feshbach resonances \cite{rmp:2010:grimm}
are preferable since 
 the life-time of the quasi-bound state in such resonances 
 will be appreciable for the  lasers to excite bound-bound transitions.  
For  $\mid \chi \rangle \leftrightarrow \mid b_n \rangle$  
bound-bound laser coupling $\Omega_n$ to be significant , one needs to  
choose the excited bound state $\mid b_n \rangle$  such that  
its outer turning point lies within an intermediate separation $r$ ($20 \le r \le 30$), 
since the wavefunction of $\mid \chi \rangle$ is usually peaked in this range. 
 It is possible to find such excited bound
states of alkali dimers having outer turning points at such intermediate separations,  and 
such bound state are accessible by PA 
transitions as demonstrated in a number of experiments. 

As an example, let us consider narrow Feshbach resonance in ultracold $^{87}$Rb atomic gas  
near magnetic field strength 
1007.4 G \cite{pra68:2003:volz,
pra70:2004:durr} in order to realize  BIC in cold collisions. 
This resonance is characterized by the parameters: zero crossing 
width $\Delta = 0.21 $ G, background scattering length $a_{bg} = 100.5 a_0$, the difference between magnetic moments 
of the closed-channel bound state $\mid b_c $ and the two free atoms is $\delta \mu = 2.79 \mu_B$, where $\mu_B$ 
is Bohr magnetron. This means the Feshbach resonance linewidth $\Gamma_f = k a_{bg} \Delta \delta \mu$ where $k$ is 
the collision wave number related to the collision energy $E = \hbar^2 k^2 /(2 \mu)$ with $\mu$ being the reduced mass
of the two atoms. For $E = 50 $ nK,   
$\Gamma_f  \sim 10 $ kHz. The values of parameters $g_1$ and $g_2$ used in Figs. 2 and 3 would
correspond to stimulated line widths of the order of 10 or 100 kHz. From 
the positions of the minimum in Figs. 2 and 3, it may be noted  that BIC will occur 
at sub-$\mu$K energy, requiring  Bose-Einstein condensate of $^{87}$Rb atoms in order to realize
a BIC near this particular  Feshbach resonance. However, 
the theoretical results depicted in Figs. 2 and 3 can fit into several other alkali 
atoms for which a condensate is not essential. Moreover, different parameter 
regimes  can be used for different alkali systems. 
In short, our model provides a vast range 
of parameter space with well defined relationship among
the various parameters for searching for BIC in cold collisions.

\subsection{Detection of BIC via photoassociative ionization spectroscopy}

BIC can also  be 
detected by photoassociative ionization spectroscopy. 
In BIC scheme of Fig.1, a third laser $L_3$ can be applied to excite molecular auto-ionization transitions 
$\mid A \rangle_{{\rm BIC}} \rightarrow 
\mid b_3 \rangle $, where $\mid b_3 \rangle$ represents a bound state in an excited potential that 
asymptotically corresponds to two atoms in $P + P$ electronic states.  Since molecular state $\mid b_3 \rangle$ 
is made of two doubly excited atoms, 
it can autoionize to produce molecular ion. Since $\mid b_1 \rangle$ and $\mid b_2$ can be chosen to be energetically 
close,  $L_{3}$ can couple both of them to $\mid b_3 \rangle$. Therefore, 
we can construct a photoassociative ionization (PAI) interaction operator 
\bea 
\hat{V}_{PAI}(t) = \Omega_{31} e^{- i (\omega_{L_3} - \omega_{31})t} 
\mid b_3 \rangle \langle b_1 \mid + \Omega_{32} e^{- i (\omega_{L_3} - \omega_{32})t} \mid b_3 \rangle 
\langle b_2 \mid + {\rm C.c}
\eea
where $\Omega_{31}$ and $\Omega_{32}$ are the Rabi frequencies for the transitions 
$\mid b_1 \rangle \rightarrow \mid b_3 \rangle$ and $\mid b_2 \rangle \rightarrow \mid b_3 \rangle$, respectively; 
$\omega_{L_3}$ is the frequency of $L_3$ laser, and $\omega_{3 n}$ is the transition frequency for the transition 
$\mid b_1 \rangle \leftrightarrow \mid b_3 \rangle $. The PAI spectrum is given by 
\bea 
S_{PAI}(\omega_{L_3}) =  \left |  \int_{0}^{\infty} d \tau   \int d E e^{- i E \tau /\hbar - \gamma \tau/2}   
  \langle b_3 \mid \hat{V}_{PAI} (\tau) \mid E+ \rangle  \right |^2 
\eea 
where $\gamma$ is the non-radiative autoionizing line width of $\mid b_3 \rangle$.  Using the eigenstates $\mid \lambda_i \rangle $ ($i = 1,2,3$) of $H_{eff}$, one can  
employ  the identity operator $\sum_{i} \mid \lambda_i \rangle \langle \lambda_i \mid $ to express $ \langle b_n \mid E + \rangle $ 
in terms of  $\mid \lambda_i \rangle $ basis 
 \bea
 \langle b_n \mid E + \rangle &=& \sum_{i=1}^{3} 
 \frac{\langle b_n \mid \lambda_i \rangle \langle \lambda_i \mid V \mid E \rangle_{{\rm bare}}
}{E - \lambda_i \hbar \Gamma_f/2 + i \epsilon }  
\eea
When BIC conditions for $q_{1f} = q_{2f} = q_f $ are fulfilled, two of the $\langle \lambda_i$ are bound states in continuum,
of which one is A-type and the other is B-type. For $g_1 >\!>1$ and $g_2 >\!>1$ , the probability amplitudes of 
$\mid b_1 \rangle $ and and $\mid b_2 \rangle $ in B-type BIC will be very small. Now, since the operator $\hat{V}_{PAI}$ 
couples $\mid b_1 \rangle $ and and $\mid b_2 \rangle $  only to $\mid b_3 \rangle$, it is expected that, 
under the conditions $g_1 >\!>1$ and $g_2 >\!>1$, the laser $L_3$ will predominantly couple A-type BIC to $\mid b_3 \rangle$. 
Thus PAI spectrum can be approximated as 
\bea 
S_{PAI} &\simeq& \left | \, _{{\rm BIC}} \langle A \mid V \mid E \rangle_{{\rm bare}} \right |^2 \nonumber \\
&\times& \left | \frac{ \Omega_{3 1} \sqrt{g_2}}{ (\omega_{31} - \omega^A_{{\rm BIC}} - \omega_{L_3})+ i \gamma/2 } -  
 \frac{ \Omega_{3 2} \sqrt{g_1}}{ (\omega_{32} - \omega^A_{{\rm BIC}} - \omega_{L_3})+ i \gamma/2 } \right |^2
\eea 
where $\omega_{{\rm BIC}}^A = E_A/\hbar$ is the eigenfrequency of A-type BIC. Clearly, the spectrum will show a shift 
equal to  $\omega_{{\rm BIC}}^A$. The spectral intensity will be suppressed (enhanced) depending on whether 
the quantity 
\bea 
{\rm Re} \left [  \frac{ \Omega_{3 1} \Omega_{32}^* \sqrt{ g_1 g_2}}{ 
(\omega_{31} - \omega^A_{{\rm BIC}} - \omega_{L_3} + i \gamma/2)
 (\omega_{32} - \omega^A_{{\rm BIC}} - \omega_{L_3})- i \gamma/2) } \right ]
\eea 
is positive (negative). Thus one can detect A-type BIC by PAI with a probe laser ($L_3$) 
in the presence of two PA lasers and a magnetic field under BIC conditions. 

\subsection{Controlling cold collisions with BIC}

When BIC conditions as discussed in the model and solution sections are fulfilled, 
the eigenstate with real eigenvalue (i.e., BIC) 
effectively 
becomes decoupled from the bare continuum while the optical and the magnetic  transitions between 
the continuum  and the bound states remain active. As the system parameters are being tuned very close
to the BIC conditions, the complex eigenvalue will tend to become real. The complex eigenvalue with small imaginary part  
implies  the leakage of the probability amplitude of the BIC into the continuum. This will give rise 
to a resonant structure with extremely narrow width \cite{fonda1960} in the variation of the 
scattering cross section as a function of energy. To calculate the scattering ${\mathbf T}  $-matrix, we follow the standard 
method of scattering theory based on M\o ller operators $\Omega_{\pm}$.   The dressed 
continuum $\mid E + \rangle$ 
describes outgoing scattering waves that are influenced by laser light and the magnetic field. 
The part of the  scattering 
${\mathbf T}$-matrix element that is modified by the two laser fields and the magnetic field is 
$T_{field}(E) = \, _{{\rm bare}} \langle E \mid \hat{V} \mid E+ \rangle = 
\, _{{\rm bare}} \langle E \mid \hat{V} \Omega_{+} \mid E \rangle_{{\rm bare}}$ can be written 
as 
\bea 
T_{{\rm field }}(E) =  \,  _{{\rm bare}} \langle E \mid  V (P + Q) G(z + i \epsilon) (P + Q) V \mid E \rangle_{{\rm bare}}.
\eea 
Since $ \,  _{{\rm bare}} \langle E \mid  V  Q = 0 $ and  $ Q V \mid E \rangle_{{\rm bare}} = 0 $,  
 we have $T_{{\rm field }}(E) = 
\,  _{{\rm bare}} \langle E \mid V P G P  V \mid E \rangle_{{\rm bare}}$.  
Now, using the relation   $PG(E + i \epsilon) P = (E - H_{eff} + i \epsilon )^{-1}$ 
we can express 
\bea  
T_{{\rm field }}(E) = \frac{1}{{\rm Det}[(\tilde{E} - \tilde{H}_{eff}]} \sum_{n=1}^{3} V_n^*(E)
\sum_{m=1}^{3} {\mathscr A}_{n m } V_m (E)    
\label{tmat}
\eea
where $V_n (E) = \Lambda_n(E) $, for $n = 1,2$ and $V_3 (E) = V_E$ are the
free-bound coupling constants. Assuming these coupling constants to be real, we have 
$\Lambda_n(E) = \sqrt{\Gamma_f/\pi} \sqrt{g_n}$ and $V_E = \sqrt{\Gamma_f/\pi}$. The form of the term  
${\mathscr N}_n = \sum_{m=1}^{3} {\mathscr A}_{n m } V_m (E)$ 
for each $n=1,2$  is equivalent to that of the numerator of $S_n(E)$ 
described in subsection A.
We have proved earlier that the numerator of $S_n(E)$ has a zero for a real eigenvalue of $H_{eff}$. 
For $n=3$ we have the term 
\bea 
&&{\mathscr N}_3 =  - \frac{\Gamma_f}{2 \pi} \left [ (\tilde{E} - \tilde{\delta}_2 )^2 + (\tilde{E} - \tilde{\delta}_2 )\left \{
\tilde{\delta}_2 - \tilde{\delta}_1 + q_{1 f} g_1 + q_{2 f} g_2 \right \} + q_{2 f} g_2
(\tilde{\delta}_2 - \tilde{\delta}_1)  \right ] 
\eea 
This expression shows that while the  term  ${\mathscr N}_3 $ will not,  in general,  
vanish for a real eigenvalue of $\hat{H}_{eff}$ while the others two terms ${\mathscr N}_1 $ and 
${\mathscr N}_2 $ will do. ${\mathscr N}_3$ will vanish for a real eigenvalue when the commutative condition 
is fulfilled. This means that for A-type or B-type BIC as discussed earlier, all three terms ${\mathscr N}_n$ ($n=1,2,3$) in the 
numerator will lead to Fano-like structures with a minimum and spike-like maximum, and as the energy will approach towards 
the minimum the spike will become narrower as in PA absorption spectrum discussed earlier. 

Ideally speaking, exactly at  BIC there will be no  outgoing scattered waves. The reason is obvious - a bound state with 
infinite lifetime can not give rise  to any outgoing wave. So, to detect a signature of BIC via scattering resonances, 
the BIC should have a small but finite width meaning the eigenvalue should have small imaginary part. 
The above analysis implies that, 
when the system parameters are tuned closed to a BIC,
the first two terms ${\mathscr N}_1 $ and ${\mathscr N}_2 $ that describe the  contributions from the 
two excited bound states will cause Fano-like structure in the $T$-matrix element. Further, when the commutativity condition 
is fulfilled or nearly fulfilled, A-type or B-type BIC will show up as prominent Fano-like 
resonances since all three terms in the numerator of Eq. (\ref{tmat}) will contribute to the resonance structures. Thus,
BIC in cold atoms can be utilized for 
 narrowing  magnetic or optical Feshbach resonances or enhancing the lifetime of the resonances. Thus, creating a BIC 
 in cold atoms with lasers, it is possible to manipulate resonant interactions between the atoms.

\subsection{Efficient production of Feshbach molecules using BIC} 

Here we discuss how BIC can help in efficient production of Feshbach molecules \cite{rmp:2006:kohler}
by stimulated 
radio-frequency spectroscopy. 
Note that, for  $g_1 >\!> 1 $ and $g_2 >\!> 1$, the amplitude coefficient of $\mid b_c \rangle$ in B-type BIC is 
much greater than those of the two excited bound states. 
This means that when the two stimulated linewidths $\Gamma_1$ and $\Gamma_2$ are much 
greater than the Feshbach resonance linewidth $\Gamma_f$, by tuning the two detuning parameters to fulfill the BIC condition 
$\delta_1 = \delta_2 = q (\Gamma_1 + \Gamma_2 - \Gamma_f)$, one can prepare a B-type BIC with large probability amplitude 
for the 
closed channel bound state, which then can be converted into a Feshbach molecule by stimulated 
radio-frequency spectroscopy. 
The efficiency of the commonly used method of magnetic field sweep for conversion of pairs of bosonic atoms into Feshbach molecules
can not usually go beyond 30\%. In contrast, the efficiency of  BIC-assisted Feshabch molecule formation
can be close to unity. For experimental realization of BIC-assisted Feshbach molecule formation, 
one can use ultracold Na atoms in the parameter 
regime of the experiment by  Inouye {\it et al.}  \cite{nature:1998:inouye} and Xu {\it et al.} 
\cite{prl:2003:xu}. 

It is thus possible to suppress the atom loss in magnetic Feshbach resonance (MFR) in a Bose-Einstein condensates  
by creating BIC with two lasers.  This loss 
 occurs primarily due to the disintegration of quasi-bound states into non-condensate atoms 
that can escape from the trap. To account for the  loss,
van Abeelen and Verhaar \cite{prl83:1999:verhaar} have introduced a ``local'' lifetime of 
quasibound state $\mid \chi \rangle$ due to its  exchange coupling  to the incoming open channel 
at an intermediate separation. For sodium condensate,  this coupling occurs 
at  $r \le 24$ and the local lifetime $\tau_0 = \frac{1}{\gamma_0} = 1.4 \mu$s \cite{prl83:1999:verhaar}, where $\gamma_0$ 
is the width due to the coupling. By choosing the bound states $\mid b_1 \rangle$ 
and $\mid b_2 \rangle$ having outer turning points near 24 $a_0$ and making the 
bound-bound laser couplings $\Omega_1$ and $\Omega_2$ greater 
than $\gamma_0$, one can expect to  suppress the atom loss to some extent. But, substantial suppression 
of atom loss will result when the collision energy is tuned closed to the energy of a BIC. 
As we have analyzed earlier, scattering $T$-matrix element shows a minimum 
when energy becomes equal to the energy of one of the two bound states in continuum. 
Since the width $\gamma_0$ is given by the energy derivative of the scattering phase shift at the energy 
of quasi-bound state \cite{book:coltheo:joachain}, in the context of our model $\gamma_0$ will correspond to the energy 
derivative of the phase shift at the minimum point. Thus,  our model provides $\gamma_0 \simeq 0$ and so atom loss in 
magnetic Feshbach resonances of Bose-Einstein condensates can be largely suppressed. Experimental realization 
of the  effect of the suppression  of atoms loss in MFR in sodium BEC or in ultracold sodium gas is possible. Because, 
photoaasociation of sodium atoms 
into  relatively shorter-ranged  (outer turning 
points near $r \sim 24 a_0$ ) bound states in $1_g$ potential have been experimentally demonstrated 
\cite{prl71:1993:lett,jcp101:1994:phillips, prl73:1994:napolitano}
and used to create  light force in PA \cite{pra75:2007:gomez} and to manipulate higher partial-wave 
interactions \cite{prl:2009:deb} via optical 
optical Feshbach resonance (OFR) \cite{prl:1996:fedichev}.

MFR induced Atom loss in BEC  
is more severe partly due to bosonic stimulation unlike  that  in degenerate Fermi gases.
Feshbach molecular dimers formed of 
fermionic atoms are found to be more stable \cite{prl91:2003:shlyapnikov,prl93:2004:shlyapnikov} due to Pauli blocking. We therefore predict that the formation of fermionic 
Feshbach molecules by stimulated radio-frequency spectroscopy using BIC will be quite efficient.

\subsection{Two bound states coupled to the continuum}

In our model we have so far considered three bound states coupled to the continuum, with one being quasibound 
state embedded in the ground-state continuum and the two others being excited molecular states. Naturally, question 
arises as to what happens to the BIC if coupling to one of the bound states is turned off. Let us first consider that 
one of the lasers, say $L_2$ is absent, that is $g_2 = 0$. Then the effective Hamiltonian reduces to a $2 \times 2$ 
matrix with second row and second column of the matrix being removed. Writing the resulting 
$2 \times 2$  effective Hamiltonian in the form ${\mathbf A}_1 + i {\mathbf B}_1$, the matrix ${\mathbf B}_1$ has 
one eigenvalue equal to zero and the other one equal to $- (\hbar \Gamma_f/2) (g_1 + 1)$. Taking
$E_c \simeq 0$,  
the condition for the existence 
of a real eigenvalue is $\tilde{\delta}_1 = q_{1 f} g_1 - q_1$ which is also the condition for the commutativity between 
${\mathbf A}_1$ and ${\mathbf B}_1$. The real eigenvalue is -$q_1$ and the corresponding eigenvector is 
\bea 
\mid \psi \rangle_{{\rm BIC }} = \frac{1}{\sqrt{\Gamma_1 + \Gamma_f }} \left [ \sqrt{\Gamma_f} \mid b_1 \rangle - \sqrt{\Gamma_1} \mid \chi \rangle \right ]
\eea 
Now, for $E_c \ne 0$, by measuring the energy 
from $E_c$, we recover the standard result for condition of the occurrence of  Fano minimum 
\bea 
\frac{E - E_c}{\hbar \Gamma_f/2} = - q_{1 f} 
\eea 
at which population trapping occurs in the state $\mid \psi \rangle_{{\rm BIC }}$ . This should be manifested as 
a prominent minimum in the scattering cross section or PA rate  versus energy plot \cite{jpb:2009:deb} as in 
the case of 3 bound states coupled to continuum as discussed above.
In fact, a few years back,  two  
experiments \cite{junker:prl:2008, nphys:2009:rempe} have demonstrated  
minimum in PA loss rate near the resonant value $B_0$ of the  magnetic field that induces a Feshbach resonance. 
Though, spectroscopy of photoassociative atom loss or PA loss  is an incoherent method, the spectral minimum 
observed in such incoherent spectrum  might be related to a state closely related to  
$\mid \psi \rangle_{{\rm BIC }}$.
It is expected that in coherent PA spectroscopy or in the measurement of  scattering 
cross sections near $B_0$ under the above-mentioned BIC condition, one would be able to observe  the discussed minimum 
and an ultra-narrow resonant structure as a clear signature of the occurrence of BIC. In the experiment 
of Junker {\it et al.} \cite{junker:prl:2008}, the quasi-bound state $\mid \chi \rangle$ is probably weakly 
coupled ($\Omega$ being small) to the excited bound state, since the bound-state chosen was relatively long-ranged 
ensuring stronger free-bound Franck-Condon overlap rather than bound-bound coupling. 
This means that $q_{1 f}$ should be negative \cite{jpb:2009:deb}
and so the minimum was expected to occur on the positive side of the scattering length, and indeed that was 
the case in Ref.\cite{junker:prl:2008}.   
In contrast, the experiment by Bauer {\it et al.} \cite{ nphys:2009:rempe}  used a relatively shorter
ranged excited bound state and 
the minimum (though not very prominent) occurred very close to the resonant magnetic field where $a_s \rightarrow \infty$. 
The minimum 
position shows slight shift 
towards negative side of $a_s$ as the laser is blue-detuned by about 3 MHz (the subplots of Fig.3 of Ref. 
\cite{nphys:2009:rempe} should be compared). Assuming $q_{1 f}$ to be positive, the BIC condition provides 
$\delta_1 = \omega_{b_1} - \omega_{L_1} = - q_{1 f} ( \Gamma_1 - \Gamma_f)$. Since in experiment of Ref. 
\cite{nphys:2009:rempe}, a narrow Feshbach resonance is used, and relatively strong PA laser is used,  
$( \Gamma_1 - \Gamma_f) > 0$. With blue detuning ($\omega_{L_1} > \omega_{b_1}$,
the BIC condition will only be fulfilled if $q_{ 1 f}$ 
is positive ensuring significant bound-bound coupling. 
Autler-Townes double-peaked spectral shape will arise when the the real parts 
of the two eigenvalues are not very far apart. If BIC condition is maintained more precisely, it is expected that one of the 
peaks would be very narrow and sharp and would correspond to BIC while the other would  be relatively broad due to the fact 
that the other eigenvalue being 
essentially complex.  

We next discuss the situation when the state $\mid \chi \rangle$ or the magnetic field is is absent. 
Writing the resulting $2 \times 2 $ matrix in the 
form ${\mathbf A}_2 + i {\mathbf B}_2$, one finds that  ${\mathbf B}_2$ has a zero eigenvalue and the other eigenvalue is 
equal to $- (\Gamma_1 + \Gamma_2)$. The effective Hamiltonian has a real eigenvalue when 
$\delta_1 = \delta_2$. Note that $\delta_1$ and $\delta_2$ refer to the detuing from the light-shifted bound states. 
The eigenvalue is $\delta = \delta_1 = \delta_2$ and the corresponding eigenvector is 
\bea 
\mid \phi \rangle_{{\rm BIC }} = \frac{1}{ \sqrt{\Gamma_1 + \Gamma_2 }} \left [ \sqrt{\Gamma_2} \mid b_1 \rangle - \sqrt{\Gamma_1} \mid b_2 \rangle \right ]
\eea 
which is an excited molecular dark state which has been found to be useful to make  an optical Feshbach resonance (OFR) 
more efficient \cite{pra:2014:subrata}. 

\section{conclusions} 
In conclusions,  
we have demonstrated theoretically that it is possible 
to create and manipulate  bound states in continuum in atom-atom cold collisions 
by lasers and a magnetic field, employing currently available techniques of 
photoaasociation and magnetic Feshbach resonance. Our model is composed 
of 3 bound states interacting with the continuum of scattering states 
between ground-state cold atoms. Within the framework of effective Hamiltonian methods, 
we eliminate the continuum and obtain an effective Hamiltonian. The eigenvectors of this effective Hamiltonian with 
real eigenvalues represent the bound states in continuum. 
We have provided specific conditions for the occurrence of a BIC
in the form of 
analytical expressions of the relationships between parameters of our model. We have 
derived photoassociative absorption spectrum and  scattering cross sections
that can exhibit signatures of a BIC as an ultra-narrow asymmetric peak 
near a prominent minimum. The minimum occurs exactly at the energy at which BIC occurs.
We have analyzed in some detail the possible applications of BIC in controlling cold collisions 
and efficient production of Feshbach molecules. 

The original proposal of von Neuman and Wigner to create a BIC of a particle was through destructive quantum 
interference of Schr\"{o}dinger's waves scattered by a specially designed potential so that there 
exists no outgoing waves resulting in the trapping of the particle in the continuum. The key mechanism 
for creating a BIC is the quantum interference which happens not only in scattering of waves but also 
in  different transition pathways in atomic and molecular physics. Bound states in continuum in our model 
result from the quantum interference in three possible free-bound transition pathways.  
In case of 2 bound states interacting with the continuum, the effective Hamiltonian 
yields one BIC that occurs at an energy at which Fano minimum takes place. This indicates   
that  BIC in our model does occur due to quantum interference in possible transition pathways.
In recent times, 
utilization and manipulation of quantum interference effects have been essential in demonstrating 
a number of coherent phenomena, paving the way for  emerging quantum technologies. Of late, 
quantum interference  are being considered for manipulating ultracold collisions \cite{mofr:deb,pra:thomas}. 
The realization of our proposed bound states in continuum in cold collisions will open a new perspective 
in quantum interference phenomena with cold atoms and molecules.

\appendix
{}

\section{Derivation of effective Hamiltonian}

\bea 
\hat{V} G = \hat{V} (P + Q) G = \hat{V} P G + \hat{V} Q G 
\label{vg}
\eea 
\bea 
Q G = Q G_0 + \frac{Q}{E - H_0 + i\epsilon} (\hat{V} P G + \hat{V} Q G)
\eea 
which leads to 
\bea 
Q G = \frac{1}{E - H_0 - Q\hat{V} + i\epsilon} \left ( Q + Q \hat{V} P G \right )
\label{qg}
\eea 
Substituting (\ref{qg}) in (\ref{vg}) and Eq. (6), after some algebra, we get  
\bea 
 P G P = \frac{1}{
E - H_0 - P \hat{V} P - P \hat{V} Q \frac{1}{E - H_0 - Q \hat{V} Q + i\epsilon }Q \hat{V} P}    \nonumber 
\eea 
which suggests that the effective Hamiltonian is 
\bea 
H_{\rm{eff}} = H_0 + P R P \eea  where 
\bea 
R =  \hat{V}  +  \hat{V} Q \frac{1}{E - H_0 - Q \hat{V} Q + i\epsilon } Q \hat{V} 
\eea 
Now, in the subspace of 3 bound states, we need to diagonalize $H_{\rm{eff}}$. Using Eqs. (2) and (4), we have 
\bea 
Q \hat{V} Q = 0
\eea 
and 
\begin{widetext}
\bea 
&& \hat{V} Q \frac{1}{E - H_0 - Q \hat{V} Q + i\epsilon } Q \hat{V} \nonumber \\
&=& \sum_{n,n'} \left [ \Delta_{n n'}^{\rm{shift}}(E) - i \frac{\Gamma_{n n'}(E)}{2} \right ] \mid b_n \rangle \langle b_{n'} \mid + 
\left [ \Delta_{f}^{\rm{shift}}(E) - i \frac{\Gamma_{f}(E)}{2} \right ] \mid b_c \rangle \langle b_c \mid \nonumber \\
&+& 
 \left [ \sum_{n} \left \{ \Delta_{n f}^{\rm{shift}}(E) - i \frac{\Gamma_{n f}(E)}{2} \right \} \mid n \rangle \langle b_c \mid +
 {\rm{C.c.}} \right ]
\eea 
\end{widetext}
where 
\bea 
\Delta_{n n'}^{\rm{shift}}(E) &=& {\cal P} 
\int d E' \frac{ \Lambda_n(E')  \Lambda_{n'}^*(E') }{E - E'  } \\
\nonumber \\
\hbar \Gamma_{n n'}(E) &=& 2 \pi  \Lambda_n(E)  \Lambda_{n'}^*(E) \\
\nonumber \\
\Delta_{f}^{\rm{shift}}(E) &=& {\cal P} 
\int d E' \frac{ |V_{E'}|^2 }{E - E'} \\
\nonumber \\
\hbar \Gamma_f(E) &=& 2 \pi  |V_{E'}|^2 \\
\nonumber \\
 \Delta_{n f}^{\rm{shift}}(E) &=& {\cal P} 
\int d E' \frac{ \Lambda_n(E')  V_{E'}^* }{E - E'  } \\
\nonumber \\
\hbar \Gamma_{n f}(E) &=& 2 \pi  \Lambda_n(E)  V_{E}^*
\eea 
$\Gamma_f $ is the Feshbach resonance line width. 
Using (A6) in (A4) and (A3), one can obtain the effective Hamiltonian of Eq. (7) the matrix elements
of which are given by 
\bea 
&& \langle b_n \mid H_{eff} \mid b_{n'}\rangle \nonumber \\
&=& (E_n - \hbar \omega_{L_n}) \delta_{n n'}  + \Delta_{n n'}^{\rm{shift}}(E) - i \frac{\hbar \Gamma_{n n'}(E)}{2}
\eea
\bea 
\langle b_c \mid H_{eff} \mid b_c \rangle = E_{0} +  \Delta_{f}^{\rm{shift}}(E) - i \frac{\hbar \Gamma_{f}(E)}{2}
\eea
\bea 
\langle n \mid H_{eff} \mid b_c \rangle =  
\Delta_{n f}^{\rm{shift}}(E) + \hbar \Omega_n - i \frac{\hbar \Gamma_{n f}(E)}{2}
\eea

\end{document}